# Online and Offline Analysis of Streaming Data

Sheik Hoque* and Andriy Miranskyy[†]
*[†] Department of Computer Science, Ryerson University, Toronto, Canada
* Royal Bank of Canada, Toronto, Canada
* sheik.hoque@ryerson.ca, [†] avm@ryerson.ca

*Abstract*—Online and offline analytics have been traditionally treated separately in software architecture design, and there is no existing general architecture that can support both. Our objective is to go beyond and introduce a scalable and maintainable architecture for performing online as well as offline analysis of streaming data.

In this paper, we propose a 7-layered architecture utilising microservices, publish-subscribe pattern, and persistent storage. The architecture ensures high cohesion, low coupling, and asynchronous communication between the layers, thus yielding a scalable and maintainable solution.

This design can help practitioners to engage their online and offline use cases in one single architecture, and also is of interest to academics, as it is a building block for a general architecture supporting data analysis.

## I. Introduction

With the advance of technology and rapid increase of Internet of Things [2] devices in our daily life, we are seeing a growing market of streaming analytic [1]. The types of analysis of streaming data vary depending on an application, but they can be grouped into two major categories: online and offline.

Not every use case needs all the incoming data; rather they are looking for a particular event to occur. Therefore, only a relevant subset of data should be delivered to a program dealing with a particular use case. In this paper, our **goal** is to introduce a layered architecture that 1) takes heterogeneous streaming data from various sources as input, 2) identifies a subset of these data relevant to our business use cases, and 3) performs analysis of the data to satisfy these use cases.

The architecture consists of a combination of 1) microservices [10], 2) instances of the publish-subscribe pattern [4], spread over multiple layers, and 3) persistent storage.

Conceptually, our architecture is based on data-flow architectural style [12, sec. 13.3.1], adapted to modern distributed (Cloud) systems. Microservices act as filters and publish-subscribe pattern provides mechanism for piping the data between the filters. When the size of a data record — to be passed in a message to an instance of the publish-subscribe pattern — is small, we can attach the data record directly to the message. However, if the record is large, its size may exceed the maximum allowable size of the message's "payload". In this case, we need to save the record in a persistent storage. The message would contain a pointer to the data record in the persistent storage rather than the record itself. This extends our previous work on handling streaming online workloads with small payload [7].

To the best of our knowledge, at present, there exists no literature on a generic framework that can handle different use cases for streaming analytics (even though streaming libraries are evolving for the last 15 years [3]), hence our focus on closing this gap. We provide a summary of related but complementary works below.

The idea of using microservices for processing streaming data is not new: multiple architectures were created for processing streaming data for specific use cases [5], [18], [9], [11]. All the micro-services in these architectures reside in a single layer. This approach works well for individual use cases. However, for large enterprises (where different interrelated use cases have to be handled simultaneously), it is important to partition business logic into multiple layers. Otherwise, dependencies between different microservices become cluttered, making maintenance and evolution complicated.

Our architecture tackles this issue by separating microservices into different layers. In the context of this paper, a layer is defined as a group of entities (either microservices or publish-subcribe topics) implementing similar business logic (e.g., converting or splitting the data). Our analysis shows that the 7-layered architecture gives good cohesion and coupling, which improves maintainability, testability, and evolvability of a software product (see Section II-D for details).

A number of specialised frameworks were created for offline processing of data, e.g., [16], [15], [13]. However, none of them are suited for processing streaming data. A process of transforming relational queries to customised distributed continuous query processing to provide streaming nature to a traditional database based application has been suggested in [20]. Microservices and publish-subscribe pattern were used [19] to analyse video streams; however, no generalisation of the approach was performed. A number of approaches also exist for speeding up extract-transform-load process (e.g., [6], [17]), but they are not designed to handle full analytics pipeline. Formal language was introduced [14] to optimise processing of streaming data, however the optimisation happens at the microservice/function level.

An inquisitive reader might wonder: why we cannot simply adopt a classic data-centered architectural style [12, sec. 13.3.1], given that we already have persistent storage in place? Unfortunately, the complex relations between the data feeds make maintainability and scalability of such solution burdensome.

## II. ARCHITECTURES

Business use cases dealing with analysis may vary, ranging from sentiment analysis of a product, to supply chain analysis, to prediction of a future stock price. Thus, the output from the models may vary, ranging from business intelligence report (e.g., identifying potential cases to improve a business), to automated notification to a warehouse (e.g., telling when to replenish the stock), to an order to a trading platform (e.g., directing to buy or sell a stock).

The primary objectives of the design are 1) to ensure a microservice does not have to process non relevant data 2) to ensure data are re-playable to avoid any data loss in case of failure in a component, and 3) to ensure isolation of components, hence easily pluggable. These design objectives bring a good impact on throughput and latency. Moreover, isolation eases the maintenance and regression testing efforts.

Formally, the number of layers in our architecture can be summarised using the formula $2n+1$. Odd layers perform data transformation, while even layers serve as a communication mechanism. That is, $i$-th layer will pass messages from $i-1$ layer to $i+1$ layer. We consider cases $n = 0, 3$ below. To preserve space, we give detailed description only of the $n = 3$ case and a cursory one of the $n = 0$ case.

### A. 1-layered architecture ($n = 0$)

This case represents 1-layered architecture where no communication channel is used. For each analytics model, there will be only one microservice performing data extraction, transformation, aggregation, and analysis.

This solution works better than a single monolith containing code of all models. The 1-layered architecture ensures scalability of each model individually and makes the code more comprehensible. In the case of failure of one microservice, the others can still be functional. However, this architecture leads to computational overhead as the microservices have to duplicate data extraction efforts.

### B. 7-layered architecture ($n = 3$)

To improve coupling and cohesion, we factor out (into separate layers) logic related to core data extraction, transformation, and analysis activities, namely, converting, splitting, aggregating, and modelling. The layers communicate with each other via an instance of publish-subscribe pattern for online data analysis. This is also acceptable for offline data analysis, if the data size is smaller than the maximum length constraint of publish-subscribe message or no historical data are needed. Otherwise, we have to resort to the persistent storage for data transmission between layers, while publish-subscribe service is used for asynchronous notifications. A diagram of the 7-layered architecture is given in Fig. 1. Details are provided below.

*1) Persistent Storage:* If the data records are small, then we can pass the records between layers in publish-subscribe messages, bypassing the persistent storage.

However, if the records are large or if we need to preserve them for future analysis (as was discussed in Section I), then $i$-th microservice layer may store the data in the persistent storage, publish a message (with the pointer to the data in the persistent storage) to the publish-subscribe layer $i+1$, so that the layer $i+2$ would be able to access the data record.

For example, a microservice in the third layer will save processed data to persistent storage and then emit a message to the fourth layer with the pointer to the stored data. A microservices in the fifth layer will be able to read the message, follow the pointer, and access the data in persistent storage for further processing.

The persistent storage may also be used to preserve configuration files of microservices, calibrated models, etc.

*2) Converter:* The first layer listens to $L$ data streams/feeds, converts the data to universal format, and publishes the data to topics maintained by the second layer. The second layer implements publish-subscribe pattern and hosts $L$ topics, working as a communication channel between the first and the third layer. There is a 1-to-1 relation between microservices in layer 1 and topics in layer 2.

*3) Splitter:* $L$ microservices in the third layer listen to $L$ topics of the second layer. There is a 1-to-1 mapping between topics of the second layer and microservices of the third layer. Each microservice applies business logic on a received data, categorises the data, and publishes them to one or more topics of the fourth layer. Those data that do not relate to any topic are discarded. Thus, there is a 1-to-many relation between microservices in the third layer and topics in the fourth layer. As is the case with the Converter, if the data are big, then we store the data in the persistent storage and publish notifications, with the pointer to the data, to topic(s) of the fourth layer (rather than publishing the data directly).

The fourth layer implements publish-subscribe pattern, hosting $X$ topics, and works as a data transmission channel between the third and the fifth layers. The actual value of $X$ is typically independent of the number of input streams and models and is dictated by the nature of use cases and the data.

*4) Aggregator:* The logic specifying a list of topics of interest for a given analytic model resides in the fifth layer. Microservices in this layer will listen to one or more topics in the fourth layer, receiving messages with the data (or pointers to the data) or notifications to start data aggregation (for offline processing).

A microservice in this layer may implement different triggers to send the data (or pointers to the data if the data are big) to a model in layer 7 (via a topic in layer 6). For example, it can relay every received message instantaneously to the model. Another example is to wait for a notification message telling the microservice that we reached a point in time specified by service level agreement and that it is time to start gathering available data. Yet another example is to wait for all the required types of data to be received (as specified by business logic) and then pass these data to the model.

There is one microservice per analytic model. The total number of models is denoted by $N$. Thus, there will be $N$ microservices in this layer. There is a many-to-many relation

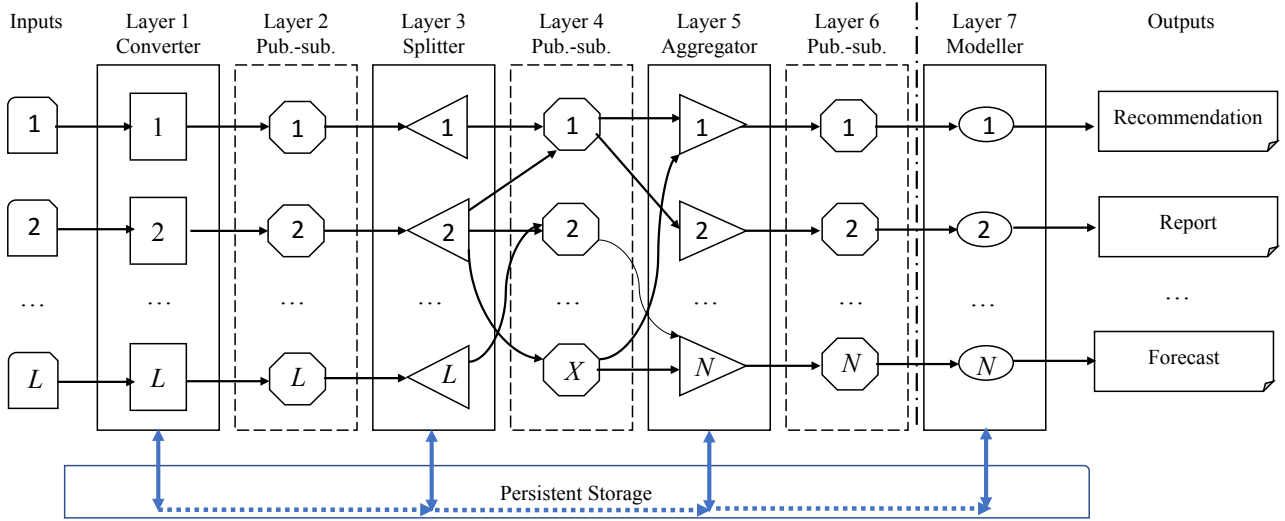

Fig. 1. A diagram of the 7-layered architecture. Dashed lines represent publish-subscribe layers. Vertical dash-dotted line separates data preparation layers from the analytics layer. Arrows denote flow of data. Blue arrows between layers 1, 3, 5, 7 and persistent storage reflect potential communication between microservices (in a given layer) and persistent storage.

between the topics in the fourth layer and the microservices in the fifth layer.

The sixth layer implements publish-subscribe pattern, one topic per model, hence $N$ topics, and works as a data transmission channel between layers 5 and 7. There is a 1-to-1 relation between microservices in layer 5 and topics in layer 6, as well as between topics in layer 6 and microservices in layer 7.

*5) Modeller:* The seventh layer implements $N$ analytic models, one microservice per model, consumes the data (or pointers to the data) from the relevant topics in the sixth layer, and executes the model. A model may also use persistent storage to read model's configuration and update model's state.

*C. Analysis of complexity*

Computational complexity for both architectures is $O(MN)$, where $M$ represents total number of messages in all input feeds and $N$ – the number of analytic models. A complete analysis comparing 1- and 7-layered architecture can be accessed in the supplementary materials [8].

The key factor of performance for 7-layered architecture is the message duplication by microservices of layer 3 into topics of layer 4. In the worst-case scenario, every message is published to every topic, setting the total number of messages in layer 4 $\gamma = MX$. However, a number of messages that are relevant to every topic is very small, thus, such scenario is not likely.

For the same relevancy reason, large number of messages from input feeds get filtered out, leading to further decrease of $\gamma$. Thus, formally $\gamma = \omega\delta$, where $\omega \in (0, X]$ is the average number of posts per message that do not get discarded, and $\delta \in [0, 1]$ is the fraction of the messages that are not discarded (on average).

We answer the question of when the 7-layered architecture produces results faster than the 1-layered one, by exploring under which conditions $T^1 > T^7$ holds (for the worst-case scenario) in [8]. The analysis shows that the inequality holds when

$$\omega\delta < (cN - c_1 - c_2 - c_3)/(c_5 N + c_6 N + c_4), \quad (1)$$

where $c$ is the highest cost of processing a message in the 1-layered architecture and $c_i$ is the highest cost of performing an operation in the $i$-th layer of the 7-layered architecture[1]. Given that $c$ is the cost of extracting, transforming and filtering data,

$$c \leq c_1 + c_3 + c_5. \quad (2)$$

This fact, in a limit as $N \to \infty$, simplifies Eq. 1 to

$$\omega\delta < (c_1 + c_3 + c_5)/(c_5 + c_6). \quad (3)$$

This gives us an estimate of the maximum value of $\omega\delta$ when 7-layered architecture will be faster than 1-layered one for a large value of $N$.

Fig. 2 suggests that we reach these limiting values for a relatively small $N$: when $N = 100$, $\omega\delta \approx 1.48$ for the former and $\omega\delta \approx 3.3$ for the latter case. That is Eq. 3 can serve as an upper bound approximation for Eq. 1 for a large $N$.

The actual values of $c_i$s would vary depending on the business use cases; thus, one has to recompute Eq. 1 and/or Eq. 3.

*D. Discussion*

This architecture provides all the benefits of microservices and asynchronous communication (via publish-subscribe software). On top of that, it has each particular data extraction and transformation feature residing in its own layer. This leads to increase of cohesion and decrease of coupling of the layers. This also improves code comprehensions (as the amount of code per microservice decreases), leading to higher maintainability.

---

[1] Note that cost of reading and writing to persistent storage is included in $c$, $c_1$, $c_3$, $c_5$, and $c_7$.

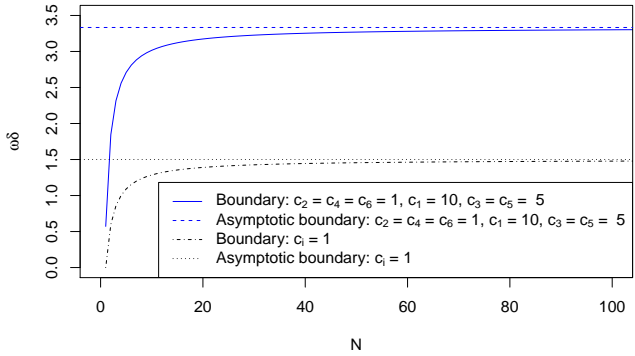

Fig. 2. Plotting $N$ vs. $\omega\delta$ in Eq. 1 for different values of $c_i$. Area under a given curve represents $N$ and $\omega\delta$ values where the 7-layered architecture is faster then the 1-layered one for a given set of $c_i$s. Horizontal lines represent the asymptotic limiting values given by Eq. 3.

High cohesion and low coupling have positive impact on operational maintenance. Cohesive code and decoupled logic help in speeding up operational activities, such as root cause analysis of failures and deployment of services.

In order to evolve a product (by adding new features) while keeping it robust, one requires to perform thorough functional and regression testing. In this architecture, the usage of microservices and isolation of logic by layer lead to smaller code base per component (microservice). This improves testability of the product, hence the cost optimisation.

To quantify the savings for our architecture, we performed complexity analysis in Section II-C. A monolith application has to process all the data, filtering them for different use cases. Our solution is layer-based: data are flowing from one layer to another and are funnelled to bring required subset of data to a particular microservice, reducing in some cases the amount of computations.

## III. Conclusion and future work

Our proposed architecture yields increased scalability and maintainability by ensuring low coupling and high cohesion of the solution. The formulas derived from formal worst-case scenario analysis convey when the 7-layered architecture is more applicable to a business use case than the 1-layered one (from computational perspective). This approach may be useful for practitioners to implement a scalable and maintainable architecture that avails their stream data for online and offline analysis. Furthermore, it is of interest to academics as a building block for a general architecture for data processing. In the future, we plan to explore applicability of this architecture to domains outside of streaming analytics realm.

## Acknowledgment

This research is funded in part by NSERC Discovery Grant No. RGPIN-2015-06075.

## References

[1] Streaming Analytics Market by Type (Solution & Services), Applications (Fraud Detection, Sales & Marketing Management, Predictive Asset Maintenance, Risk Management, Network Management, Location Intelligence, & Operations Management), Vertical, Regions - Global Forecast to 2021. Technical Report TC3451, Markets and Markets, 2016.
[2] Internet of things, Apr. 2018. Page Version ID: 835313195.
[3] M. Babazadeh and C. Pautasso. The Stream Software Connector Design Space: Frameworks and Languages for Distributed Stream Processing. In *2014 IEEE/IFIP Conf. on Softw. Architecture*, pages 1–10, Apr. 2014.
[4] K. Birman and T. Joseph. Exploiting Virtual Synchrony in Distributed Systems. In *Proceedings of the Eleventh ACM Symposium on Operating Systems Principles*, SOSP '87, pages 123–138. ACM, 1987.
[5] C. Esposito, M. Ficco, F. Palmieri, and A. Castiglione. A knowledge-based platform for Big Data analytics based on publish/subscribe services and stream processing. *Knowledge-Based Systems*, 79(Supplement C):3–17, May 2015.
[6] T. Freudenreich, P. Furtado, C. Koncilia, M. Thiele, F. Waas, and R. Wrembel. An On-Demand ELT Architecture for Real-Time BI. In *Enabling Real-Time Business Intelligence*, Lecture Notes in Business Information Processing, pages 50–59. Springer, Berlin, Heidelberg, Aug. 2012.
[7] S. Hoque and A. Miranskyy. Architecture for Analysis of Streaming Data. In *IEEE Int. Conf. on Cloud Engineering (IC2E)*, 2018. to appear.
[8] S. Hoque and A. Miranskyy. Online and Offline Analysis of Streaming Data: Suppl. Material. 2018. DOI=10.6084/m9.figshare.6127694.v1 , https://figshare.com/articles/Online_and_Offline_Analysis_of_Streaming_Data_Supplementary_Material/6127694.
[9] H. Hromic, D. L. Phuoc, M. Serrano, A. Antonic, I. P. Zarko, C. Hayes, and S. Decker. Real time analysis of sensor data for the Internet of Things by means of clustering and event processing. In *2015 IEEE Int. Conference on Communications (ICC)*, pages 685–691, June 2015.
[10] J. Lewis and M. Fowler. Microservices, 2014. https://martinfowler.com/articles/microservices.html.
[11] Y. Nakamoto, A. Yamaguchi, K. Sato, S. Honda, and H. Takada. Toward Data-Centric Software Architecture for Automotive Systems - Embedded Data Stream Processing Approach. In *2014 IEEE 11th Intl Conf on Ubiquitous Intelligence and Computing and 2014 IEEE 11th Intl Conf on Autonomic and Trusted Computing and 2014 IEEE 14th Intl Conf on Scalable Computing and Communications and Its Associated Workshops*, pages 586–589, Dec. 2014.
[12] R. S. Pressman and B. Maxim. *Software Engineering: A Practitioner's Approach*. McGraw-Hill Education, New York, NY, 8 edition, Jan. 2014.
[13] W. Shi, Y. Zhu, T. Huang, G. Sheng, Y. Lian, G. Wang, and Y. Chen. An Integrated Data Preprocessing Framework Based on Apache Spark for Fault Diagnosis of Power Grid Equipment. *Journal of Signal Processing Systems*, 86(2-3):221–236, Mar. 2017.
[14] R. Soul, M. Hirzel, B. Gedik, and R. Grimm. River: An Intermediate Language for Stream Processing. *Softw. Pract. Exper.*, 46(7):891–929, July 2016.
[15] S. Suguna, M. Vithya, and J. I. C. Eunaicy. Big data analysis in e-commerce system using HadoopMapReduce. In *2016 Int. Conf. on Inventive Computation Technologies (ICICT)*, volume 2, pages 1–6, Aug. 2016.
[16] K.-S. Tam and R. Sehgal. A Cloud Computing Framework for On-Demand Forecasting Services. In *Internet of Vehicles Technologies and Services*, Lecture Notes in Computer Science, pages 357–366. Springer, Cham, Sept. 2014.
[17] F. Waas, R. Wrembel, T. Freudenreich, M. Thiele, C. Koncilia, and P. Furtado. On-Demand ELT Architecture for Right-Time BI: Extending the Vision. *Int. J. Data Warehous. Min.*, 9(2):21–38, Apr. 2013.
[18] F. Xhafa, V. Naranjo, S. Caball, and L. Barolli. A Software Chain Approach to Big Data Stream Processing and Analytics. In *9th Int. Conf. on Complex, Intelligent, and Software Intensive Systems*, pages 179–186, 2015.
[19] W. Zhang, L. Xu, Z. Li, Q. Lu, and Y. Liu. A Deep-Intelligence Framework for Online Video Processing. *IEEE Software*, 33(2):44–51, Mar. 2016.
[20] Q. Zou, H. Wang, R. Soul, M. Hirzel, H. Andrade, B. Gedik, and K.-L. Wu. From a Stream of Relational Queries to Distributed Stream Processing. *PVLDB*, 3:1394–1405, Sept. 2010.